# A Low Cost Mars Aerocapture Technology Demonstrator

Athul Pradeepkumar Girija [1,**,**********]

[1]*School of Aeronautics and Astronautics, Purdue University, West Lafayette, IN 47907, USA*

**ABSTRACT**

The ability to launch small secondary payloads to Mars on future science missions present an exciting opportunity for demonstration of advanced technologies for planetary exploration such as aerocapture. Over the years, several aerocapture technology demonstration missions have been proposed but none could be realized, causing the technology to become dormant as it is seen as too risky and expensive to be used on a science mission. The present study proposes a low-cost Mars aerocapture demonstration concept as a secondary payload, and could pave the way for future low-cost small interplanetary orbiter missions. The proposed mission heavily leverages the mission architecture and the flight hardware of the MarCO spacecraft for a low cost mission. The 35 kg technology demonstrator would launch as an ESPA secondary payload on a future Mars mission, and would be deployed from the upper stage soon after primary spacecraft separation. The vehicle then independently cruises to Mars, where it will perform aerocapture and insert a 6U MarCO heritage CubeSat to a 200 x 2000 km orbit. The mission architecture incorporates a number of cost saving approaches, and is estimated to fit within a $30M cost cap, of which $10M is allocated for technology development and risk reduction.

___________
***********,** To whom correspondence should be addressed, E-mail: athulpg007@gmail.com

# I. INTRODUCTION

Aerocapture is a promising alternative to conventional propulsive insertion of orbiters around atmosphere-bearing bodies in the Solar System which include Venus, Mars, Titan, Uranus, and Neptune [1, 2]. Aerocapture uses atmospheric drag to slow down a spacecraft and accomplish orbit insertion from an interplanetary trajectory. Compared to propulsive insertion which requires substantial amount of propellant, aerocapture can offer significant mass and cost savings for orbiter missions [3]. Figure 1 shows an overview of the concept of operations for the aerocapture. The vehicle enters the atmosphere, flies deep enough into the atmosphere to decelerate, exits the atmosphere, performs a small periapsis raise burn, and then enters orbit. Though aerocapture has never been flown on a planetary mission, it has been studied extensively over the last four decades for applications to every atmosphere-bearing destination [4, 5, 6]. Over the years, several aerocapture technology demonstration missions have been proposed but none could be realized, causing the technology to become 'dormant' as it is seen as too risky and expensive to be used on a science mission. The present study proposes a low-cost Mars aerocapture demonstration concept as a secondary payload, and could pave the way for future low-cost small interplanetary missions.

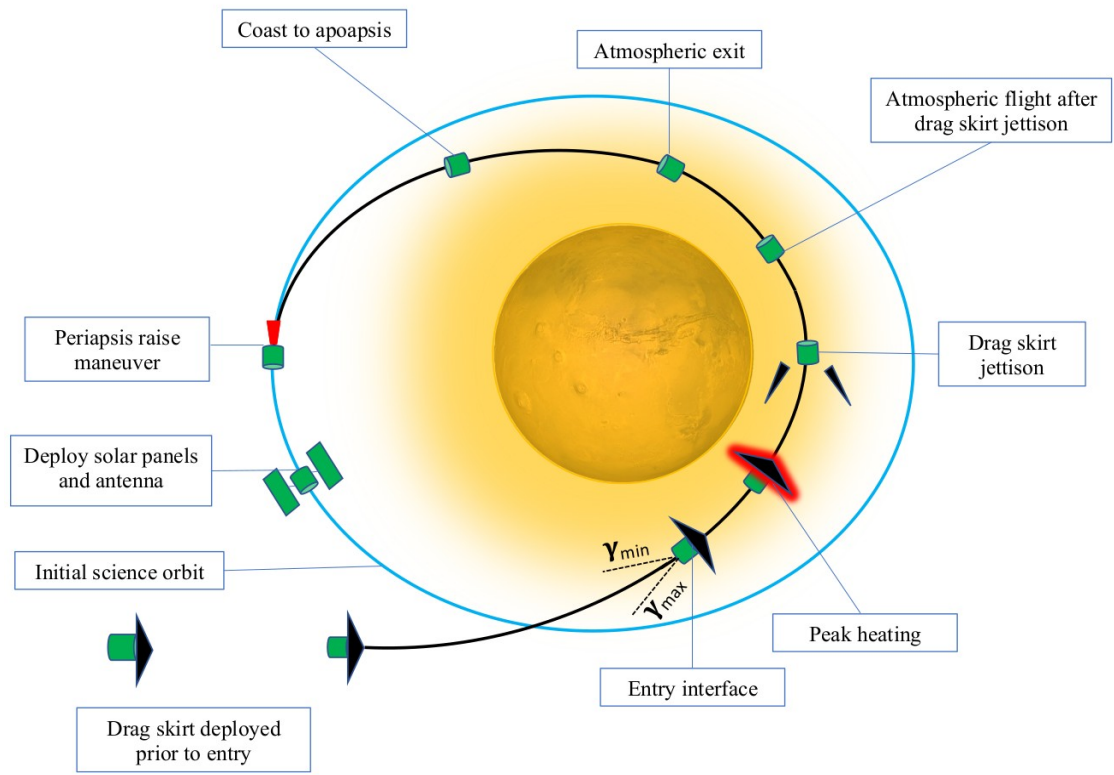

Figure 1. Concept of operations for the aerocapture mission concept.

## II. PREVIOUS TECHNOLOGY DEMONSTRATOR CONCEPTS

Starting with the Aeroassist Flight Experiment (AFE) proposal in the 1980s, several aerocapture technology demonstration missions have been proposed at Earth and Mars [7]. This study reviews the prominent efforts since 2000 towards an aerocapture technology demonstration. In 2002, Hall proposed the Aerocapture Flight Test Experiment (AFTE) as part of the New Millennium Program (NMP), a program with a focus on engineering validation of new technologies for space applications [8]. The AFTE concept in the NMP ST-7 competition proposed a low lift-to-drag (L/D) aeroshell launched to geosynchronous transfer orbit (GTO, 180 x 36,000 km) as a secondary payload. A small propulsive maneuver will target an entry trajectory for guided hypersonic flight after which the vehicle would achieve a 200 x 400 km orbit. Based on previous NMP selections, the estimated cost of the mission is in the range of $20-25M (FY 2002). NASA did not select AFTE for ST-7 although no technical show stoppers were identified. In 2003, the Aurora Program of the European Space Agency (ESA) was conceived to prepare for the human exploration of the Solar System by building up the necessary capabilities through a series of robotic missions [9]. The Aurora program proposed a Mars aerocapture demonstrator as a small precursor mission before its application to future larger missions, but program was not realized eventually. In 2005, another European initiative called the AEROFAST also proposed a Mars aerocapture demonstrator [10]. In 2013, the Japanese Aerospace Exploration Agency (JAXA) proposed a small, low-cost Mars aerocapture demonstrator using a low-L/D aeroshell but was also not flown [11]. By the early 2010s, it became increasingly clear that lifting aeroshells with their complex control systems were too difficult to fit within the small budgets for any aerocapture demonstration. In 2014, Putnam and Braun proposed the concept of drag modulation as a simpler flight system which would avoid the need for complex control systems used in lifting aeroshells, potentially reducing the cost of aerocapture vehicles by about an order of magnitude [12]. Drag modulation has since then been extensively studied for applications to small low-cost Mars and Venus missions [13, 14] . In 2019, a NASA study concluded that a technology demonstration is not required prior to using aerocapture on a science mission [1]. However, the conclusion was only based on its analysis of lifting aeroshells for large missions and did not consider drag modulation aerocapture in detail for its applications to small missions.  In 2019, Werner and Braun first proposed a low-cost aerocapture demonstration at Earth using a drag modulation vehicle [15]. The concept has since been refined by researchers to use a deployable drag skirt that lends itself to better packaging as a secondary payload [16]. The concept as shown in Figure 2 is essentially the same as that proposed by Hall in 2002, but with a simpler drag modulation flight control instead of the lifting aeroshell. In 2023, an ESA study proposed an idea for a Mars aerocapture demonstrator with a rigid drag skirt which would be launched on a dedicated Vega-C launcher [17].

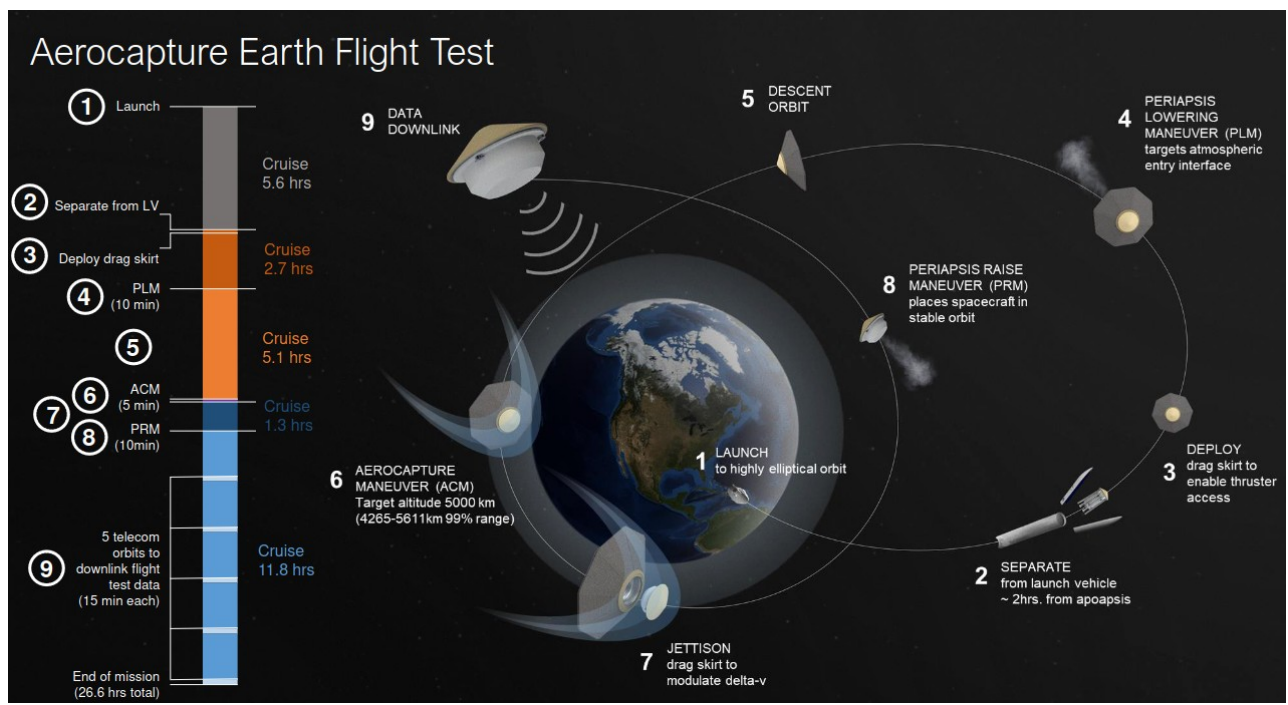

Figure 2. Concept of operations for the proposed aerocapture Earth flight test [16].

The 2023-2032 Planetary Science Decadal survey concluded that "aerocapture is considered ready for infusion and can enhance/enable a large set of missions, but that will require special incentives by NASA to be proposed and used in a science mission [18]. Aerocapture can also enable planet orbit insertion of SmallSats, launched as secondary payloads, on targets like Venus and Mars [19]. Because aerocapture is not being proposed for use in missions, it is considered a 'dormant' technology that is perceived as high risk in a mission competitive environment." The New Millenium program was discontinued in 2009, but the current Decadal Survey recommends "creating a separate technology line similar to the former New Millennium program where multiple technologies could be demonstrated in small flight missions". In 2022, a review of NASA Space Technology Mission Directorate (STMD) portfolio listed an aerocapture flight demonstration as a top priority for future exploration [20]. In 2023, the Aerocapture Demonstration Relevance Assessment Team (ADRAT) study commissioned by the NASA Science Mission Directorate concluded that an Earth aerocapture demonstration would reduce the risk to nearby destinations such as Mars and Venus with well-known atmospheres [21]. Meanwhile, NASA has announced that "all future launches would be equipped with an EELV Secondary Payload Adapter (ESPA) ring and would provide opportunities for secondary missions [22]." This presents a valuable opportunity to take advantage of the unused mass margin on a future Mars mission for an aerocapture technology demonstrator. The present study proposes a low-cost aerocapture technology demonstration to deliver a single 6U CubeSat to Mars orbit as a standalone payload within a cost cap of $30 million.

## III. PROPOSED LOW-COST MISSION CONCEPT

The proposed Mars aerocapture technology demonstrator builds upon the success of the MarCO interplanetary CubeSats, and makes extensive use of the its flight hardware and mission architecture to achieve a low-cost mission. Launched with the InSight lander as secondary payloads, the two MarCO CubeSats were the first to operate beyond Low Earth Orbit (LEO) [23]. The two spacecraft were deployed from the aft side of the upper stage after InSight seperated first as shown in Figure 3, and independently cruised to Mars. Upon arrival at Mars, the two spacecraft would target a flyby trajectory that allowed it to relay critical data from the InSight vehicle in real time during its atmospheric entry. The MarCO cubesats used innovative techniques to close the mass, power, and link budgets which are challenging to achieve for small spacecraft operating at interplanetary distances. The reported end-to-end cost of the two CubeSats combined is $18.5M, a fraction of what large interplanetary missions cost. The 6U CubeSats had a small propulsion system which could deliver about 40 m/s of $\Delta V$ for trajectory correction maneuvers en route, but had no capability to enter orbit, as it requires $\Delta V$ of several hundred m/s. The proposed mission concept uses the same architecture, and will launch as a secondary payload on a future Mars mission. The vehicle will be deployed after the primary payload separation and independently cruises to Mars, where unlike the MarCO CubeSats which only flew by Mars, it will enter the atmosphere and use drag modulation aerocapture to perform orbit insertion.

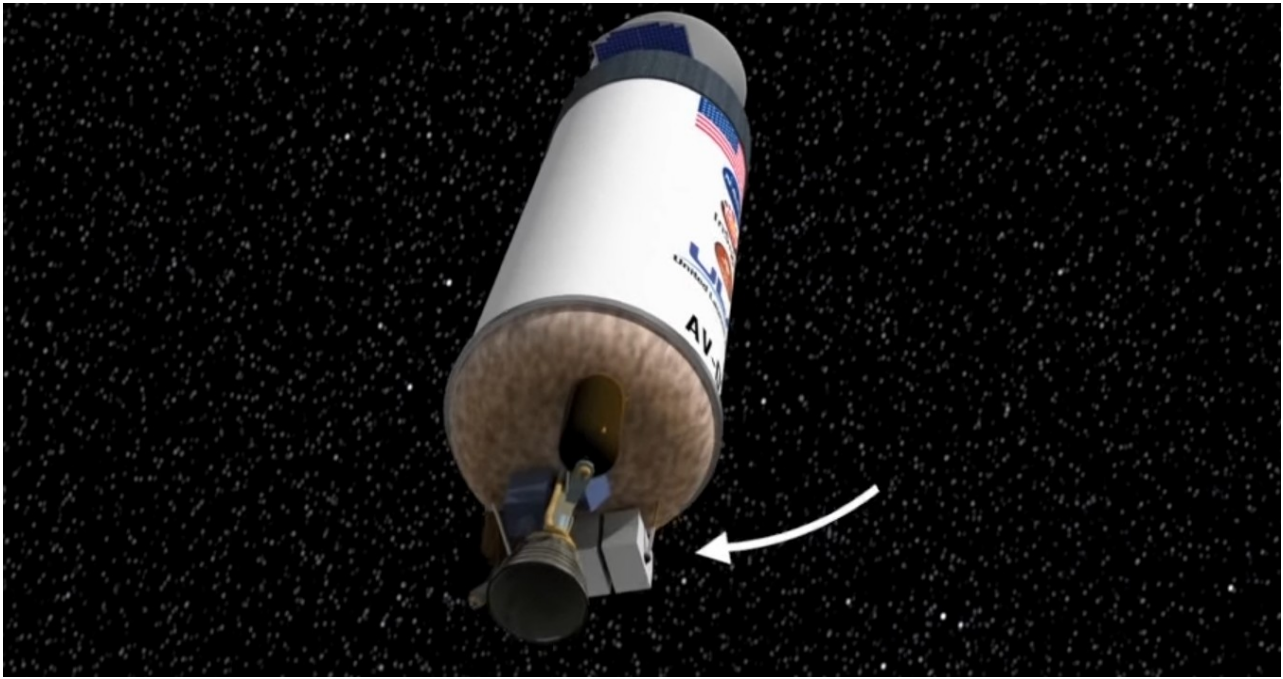

Figure 3. Location of the two MarCO cubesat deployers on the launcher upper stage.

The high-level mission objective in the study is to deliver a 6U CubeSat with MarCO heritage to a 200 x 2000 km near-polar orbit at Mars. Following MarCO, the mass of the spacecraft in orbit after aerocapture is assumed to be 14 kg. The aerocapture vehicle is based on the design presented by Austin et al. as shown in Figure 4 [24]. Assuming a typical ADEPT payload mass fraction of 0.50, and allocating 2 kg for the skirt separation mechanism, the entry system mass is estimated to be 30 kg. The diameter of the vehicle is 1.5 m when deployed, and 0.45 m when stowed following the design by Austin et al. which fits within the ESPA volume. Assuming a drag coefficient of 1.5, the ballistic coefficient of the system with the drag skirt deployed is 11.3 kg/m$^2$. Assuming a 45 deg. sphere-cone nose cap with Cd=1.5, the ballistic coefficient with the drag skirt jettisoned is 62.9 kg/m$^2$. This gives the vehicle a ballistic coefficient ratio of 5.56, which will be shown to provide more than enough control authority for aerocapture at Mars.

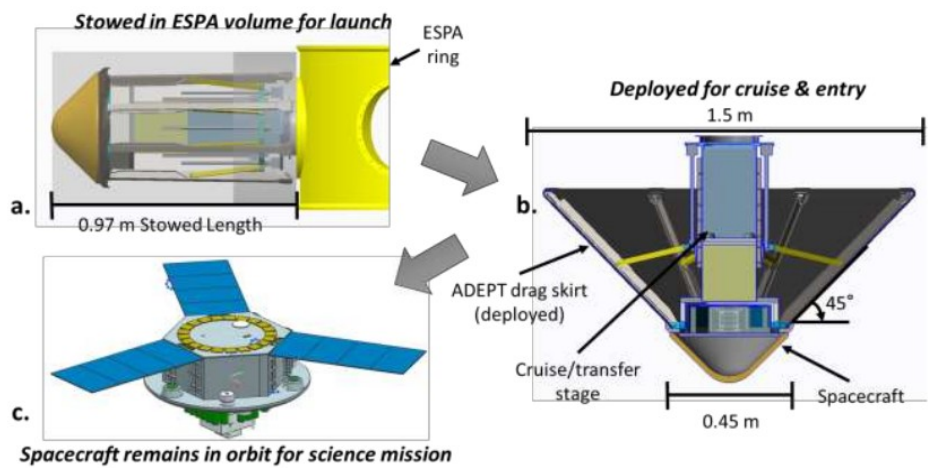

Figure 4. Stowed and deployed configurations of the ADEPT aerocapture vehicle [24].

## IV. AEROCAPTURE TRAJECTORY

The interplanetary arrival trajectory of the Mars 2020 mission is used as a reference, with the approach trajectory targeting an entry near the north pole to achieve a polar orbit. The atmospheric entry state is given in Table 1.

Table 1. Mars aerocapture atmospheric entry state

| Parameter | Value |
| --- | --- |
| Altitude, km | 120 |
| Longitude, deg | 25.06 |
| Latitude, deg | 62.35 |
| Speed, km/s | 5.6053 |
| Heading, deg | 91.2 |
| Flight-path angle, deg | -8.40 |

7Figure 5 shows the approach trajectory of the aerocapture vehicle over the Martian north pole.

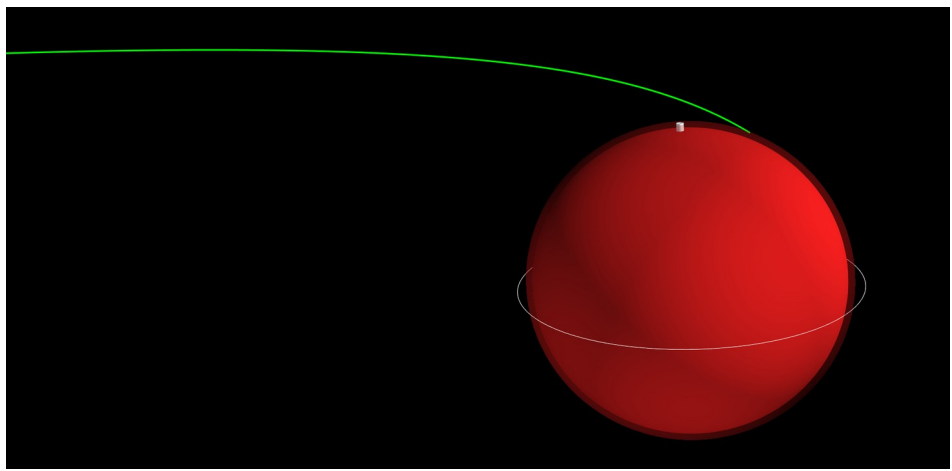

Figure 5. Approach trajectory of the Mars aerocapture vehicle. The white ring is used to indicate the equatorial plane, and the dash indicates the North Pole.

The open-source Aerocapture Mission Analysis Tool (AMAT) is used to compute the aerocapture corridor [25]. The aerocapture entry corridor is [-8.99, -7.93] deg, which provides approximately 1 deg. of Theoretical Corridor Width (TCW). The chosen entry flight-path angle of -8.40 is near the middle of the corridor. Figure 6 shows the nominal trajectory of the Mars aerocapture vehicle, with the discontinuity in deceleration indicating the jettison event.

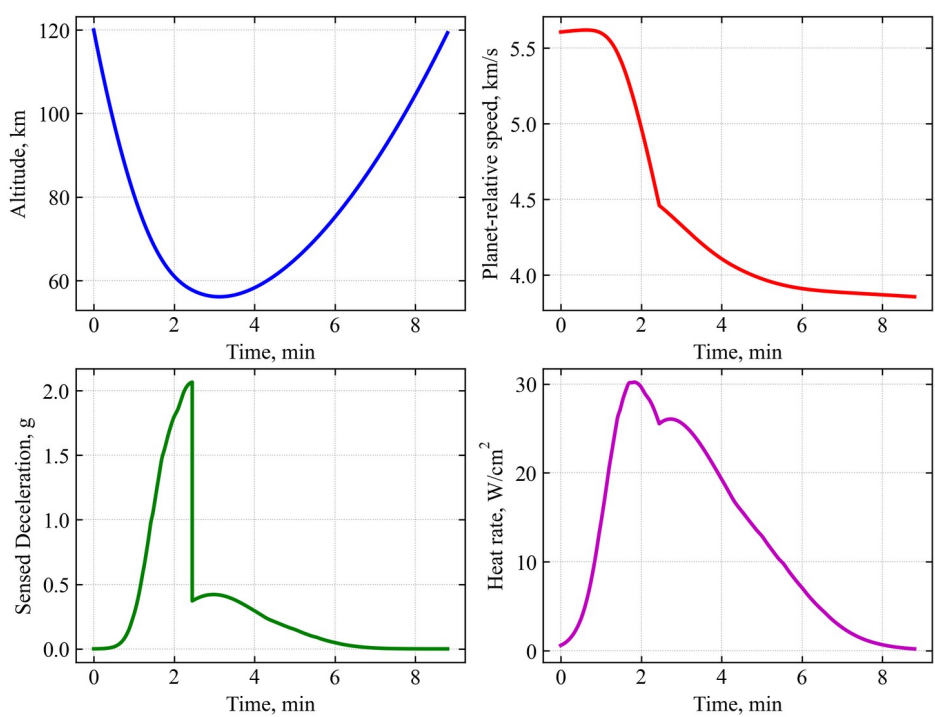

Figure 6. Nominal Mars aerocapture trajectory evolution.

Figures 7 and 8 show the sample results of Monte Carlo simulations used to assess the effect of uncertainty in the entry-flight path angle (±0.2 deg, 3σ) and atmospheric uncertainties from MarsGRAM. The results indicate reasonable orbit targeting accuracy for a technology demonstration even with a simple guidance scheme. More advanced guidance schemes have been proposed which can provide much better orbit accuracy. The deceleration loads are quite modest, just over 2g. The peak heat rate is 27-33 W/cm$^2$, which is well within the capability of ADEPT, which has been tested at over 200 W/cm$^2$, and the total heat load is in the range of 4.5-7 kJ/cm$^2$. The periapsis raise maneuver (PRM) ΔV is 30 m/s, well within the capability of a small cold-gas thruster used on the MarCO mission.

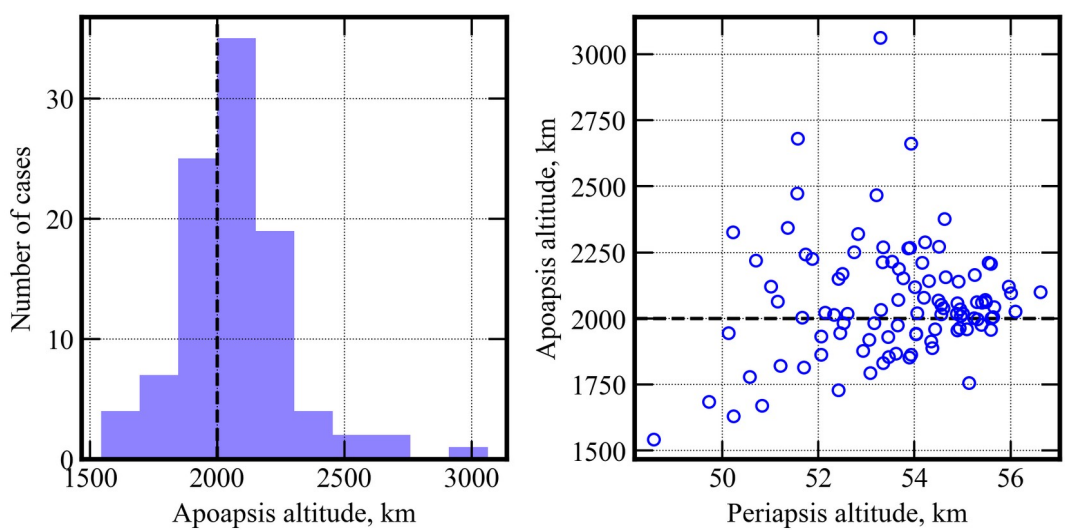

Figure 7.  Sample of post-aerocapture orbit distribution (before periapsis raise).

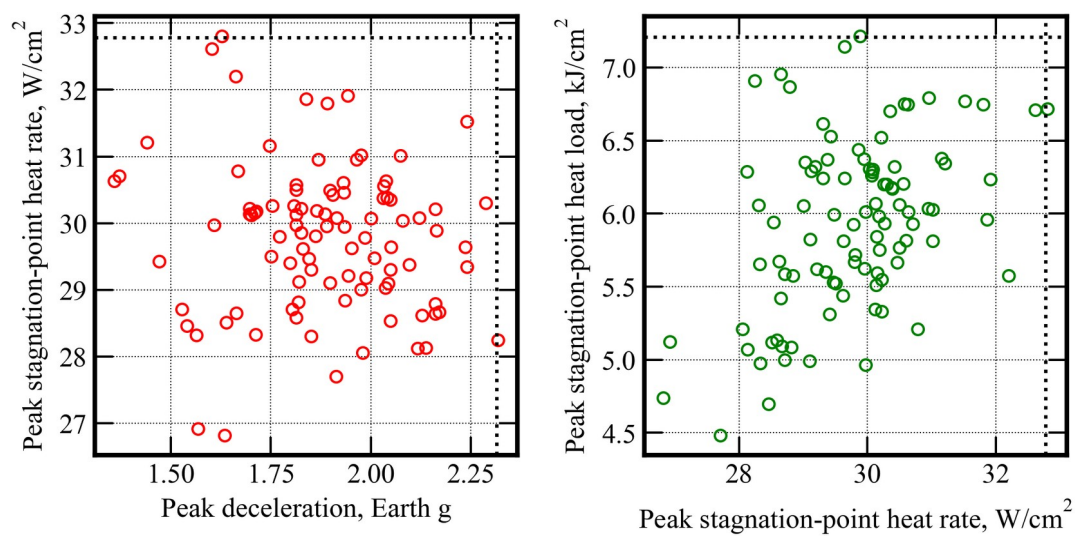

Figure 8.  Sample of deceleration, peak heat rate, and total heat load.

## V. COST SAVING APPROACH

The proposed concept incorporates several cost saving approaches compared to earlier demonstration concepts [26]. 1) Compared to previous concepts using a lifting aeroshell which requires complex control systems, the drag modulation vehicle incorporates a simpler flight control system with the only active element during the atmospheric flight being the drag skirt jettison. 2) Compared to previously proposed lifting aeroshell concepts which weighed 200-250 kg, the drag modulation vehicle weighs only 30 kg, nearly an order of magnitude less, which translates into cost savings. 3) Compared to previous concepts which would require a dedicated launch vehicle, the proposed concept uses the unused mass margin on a future Mars mission which NASA has committed to make available for secondary ESPA payloads. This essentially means the launch opportunity is free and there are no launch costs. Mars missions are quite frequent, and if one mission cannot accommodate the secondary payload, it can wait for the next opportunity. 4) The orbiter spacecraft fully leverages the technology development that was achieved by the MarCO mission for small missions operating in interplanetary space. 5) As a technology demonstration mission, the spacecraft carries no science instruments to minimize mass, and will nominally operate only for a few days in orbit to reduce the mission cost to the fullest extent possible. A small camera such as that carried by MarCO, may be optionally added for public outreach activities. 6) Mars presents most benign entry conditions for aerocapture in the inner Solar System in terms of deceleration and aerothermal environment, for which the ADEPT system is already well qualified and minimizes additional technology development cost. In addition, Mars offers a higher corridor width for aerocapture than Earth or Venus, thus increasing the available control authority. 7) The concept adheres to the principle of `do no harm' to the primary mission. The primary vehicle will be deployed first from the launcher, and the aerocapture vehicle will be deployed afterwards from the upper stage. The vehicle deploys the ADEPT drag skirt, and independently cruises to Mars, separated by a sufficient distance from the primary spacecraft and poses no risk to the primary mission.

Cost estimation of the proposed demonstration mission is a challenging task, but it is possible to provide some estimates. For reference, the 14 kg MarCO CubeSat each cost approximately $10M. The proposed vehicle is nearly twice the mass at 35 kg (30 kg entry system + 5 kg cruise stage). Using mass as a rough proxy for the cost of the very similar MarCO mission, the estimated mission cost is $20M. Edwards et al. performed a study assessing the cost drivers for small missions, and presented cost data for a chemical propulsion system vehicle as a function of $\Delta V$ and payload mass, which is shown in Figure 9 [27]. The proposed spacecraft would be equipped with two cold gas thrusters, one on the cruise stage for trajectory correction maneuvers ($\Delta V = 40$ m/s), and one on the orbiter capable of $\Delta V = 50$ m/s. The orbiter periapsis raise maneuver requires 30 m/s, and includes a 20 m/s safety margin. Thus the total propuslive ΔV capability of the spacecraft is 100 m/s. Fig. 9 shows the estimated mission cost for a mission with 100 m/s ΔV and payload mass in the range of 2-5 kg is $20M, in agreement with the rough estimate. Of the allowed budget if $30M, the remaining $10M is allocated for technology development and risk reduction activities.

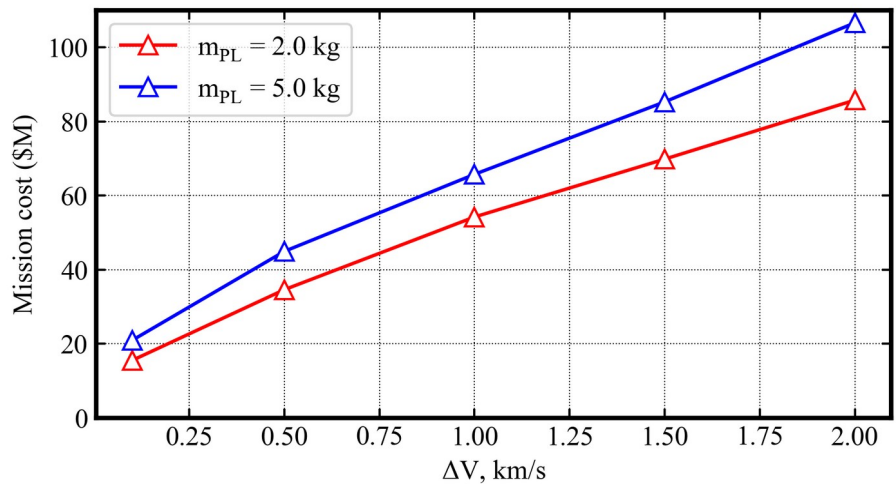

Figure 9. Estimated mission cost as a function of ΔV and payload mass.

## VI. RISK FACTORS

As with any technology demonstration mission, there are a few risks associated with the proposed technology demonstration concept. 1) The main and highest risk item is that the ADEPT drag skirt jettison system needs additional development and testing before a flight mission to bring into Technology Readiness Level (TRL) 5. This includes CFD modeling of the jettison event, ballistic range tests, and a flight test with a sounding rocket to demonstrate the jettison event during supersonic/hypersonic flight. Most of the $10M allocated for technology development would go towards this effort towards risk reduction of the jettison event. 2) There is a schedule risk that the project cannot complete all required technology development in time to meet the launch deadline of a primary mission. However, the impact is low as Mars missions are frequent enough and an alternative launch opportunity can be used. 3) There is a risk of the spacecraft impacting the surface or not getting captured. The probability of this event is low, as can be demonstrated using Monte Carlo simulations. 4) There is a risk of the thermal protection system failure. This risk is low, as ADEPT has already been tested under much more severe conditions in ground tests. 5) There is a risk of the periapsis raise maneuver failing, and the spacecraft enters the atmosphere in the next orbit after aerocapture. The risk is medium to high, but can be addressed. Overall, all the medium to high risk factors can be addressed through technology development efforts within 2-3 years, and within a reasonable budget.

## VII. IMPLICATIONS FOR FUTURE MISSIONS

Aerocapture has been studied for over four decades, with several technology demonstration concepts proposed along the way to bring the technology to flight status [28]. However, the failure to materialize a demonstration has led to the technology not being used for flight missions, even though all the required technologies exist. A successful low-cost demonstration either at Earth or at Mars as proposed in this study will establish flight heritage for aerocapture and allow its adoption for future science missions. In the short term, it will enable a new class of small and frequent interplanetary missions to Mars and Venus, including SmallSat constellations. In the long term, it will pave the way for the use of aerocapture for more ambitious missions such as multi-element Venus flagship [29], Venus atmospheric sample return [30], Titan orbiter [31], Uranus orbiter and probe [32, 33, 34], and Neptune exploration missions [35].

## VII. CONCLUSIONS

The ability to launch small secondary payloads to Mars on future science missions present an exciting opportunity for demonstration of advanced technologies for planetary exploration such as aerocapture. The present study proposed a low-cost Mars aerocapture technology demonstrator mission concept. The proposed mission heavily leverages the mission architecture and the flight hardware of the MarCO CubeSat spacecraft for a low cost mission. The 35 kg technology demonstrator would launch as an ESPA secondary payload on a future Mars mission, and would be deployed from the upper stage soon after primary spacecraft separation. The vehicle then deploys the ADEPT drag skirt and independently cruises to Mars, where it will perform aerocapture and insert a 6U MarCO heritage CubeSat to a 200 x 2000 km orbit. The peak heat rate encountered by the vehicle is about 30 W/cm$^2$, which is well within the tested capability of the ADEPT system. The periapsis raise maneuver $\Delta V$ is 30 m/s, well within the capability of cold gas thrusters used by MarCO. The mission architecture incorporates a number of cost saving approaches, and is estimated to fit within a $30M cost cap, of which $10M is allocated for technology development and risk reduction. The primary risk reduction would be the demonstration of stable hypersonic flight of ADEPT and the clean drag skirt jettison event which can be accomplished in a sounding rocket flight test. The success of the proposed mission concept will establish flight heritage for aerocapture, enabling a new paradigm of small low-cost interplanetary missions in the short term, and will pave the way for more ambitious outer planet aerocapture Flagship missions in the long term.

## DATA AVAILABILITY

All the data can be reproduced using the open-source Aerocapture Mission Analysis Tool (AMAT) v2.2.22. The data will be made available by the author upon reasonable request.